\documentclass[a4paper,10pt]{article}

\usepackage{comment}
\usepackage{float}
\usepackage{amsmath}
\usepackage[pdftex]{color,graphicx,hyperref}
\usepackage{amsmath,amsfonts,amssymb}
\usepackage[margin=1in]{geometry}
\usepackage{setspace}
\usepackage{booktabs}
\usepackage[labelfont=bf,labelsep=period]{caption}
\usepackage{subcaption}
\usepackage{cite}
\usepackage[utf8]{inputenc}
\usepackage[affil-it]{authblk}

\newcommand{\eref}[1]{Eq.~(\ref{#1})}
\newcommand{\esref}[2]{Eqs.~(\ref{#1})-(\ref{#2})}

\newcommand{\fref}[1]{Fig.~\ref{#1}}

\newcommand{\srefsi}[1]{SI Section~S{#1}}

\providecommand{\keywords}[1]{{\small {\bf Keywords---} #1}}

\title{Disentangling degree and tie strength heterogeneity in egocentric social networks}
\author[1*]{Sara Heydari}
\author[2,3,1,4]{Gerardo Iñiguez}
\author[3,5]{János Kertész}
\author[1*]{Jari Saramäki}

\affil[1]{\small{Department of Computer Science, Aalto University School of Science, 00076 Aalto, Finland}}
\affil[2]{\small{Faculty of Information Technology and Communication Sciences, Tampere University, 33720 Tampere, Finland}}
\affil[3]{\small{Department of Network and Data Science, Central European University, 1100 Vienna, Austria}}
\affil[4]{\small{Centro de Ciencias de la Complejidad, Universidad Nacional Auton\'{o}ma de M\'{e}xico, 04510 Ciudad de México, Mexico}}
\affil[5]{\small{Complexity Science Hub, 1080 Vienna, Austria}}
\affil[*]{\small{Corresponding authors' email: sara.heydari@aalto.fi, jari.saramaki@aalto.fi}}
\date{}

\begin{document}

\maketitle

\begin{abstract}
The structure of personal networks reflects how we organise and maintain social relationships. 
The distribution of tie strengths in personal networks is heterogeneous, with a few close, emotionally intense relationships and a larger number of weaker ties.
Recent results indicate this feature is universal across communication channels. 
Within this general pattern, there is a substantial and persistent inter-individual variation that is also similarly distributed among channels. The reason for the observed universality is yet unclear---one possibility is that people's traits determine their personal network features on any channel.
To address this hypothesis, we need to compare an individual's personal networks across channels, which is a non-trivial task: while we are interested in measuring the differences in tie strength heterogeneity, personal network size is also expected to vary a lot across channels. Therefore, for any measure that compares personal networks, one needs to understand the sensitivity with respect to network size.
Here, we study different measures of personal network similarity and show that a recently introduced alter-preferentiality parameter and the Gini coefficient are equally suitable measures for tie strength heterogeneity, as they are fairly insensitive to differences in network size.
With these measures, we show that the earlier observed individual-level persistence of personal network structure cannot be attributed to network size stability alone, but that the tie strength heterogeneity is persistent too.
We also demonstrate the effectiveness of the two measures on multichannel data, where tie strength heterogeneity in personal networks is seen to moderately correlate for the same users across two communication channels (calls and text messages).
\end{abstract}


\section{Introduction}

Social relationships play a fundamentally important role in our lives. On an individual level, they constitute a person's social capital\cite{Putnam2000Bowling}: they offer social support and engagement, transmit social influence, and provide access to resources and commodities\cite{Berkman2000}. At the system level, they keep society connected and functional\cite{Coleman1990}.
Social relationships vary in their purposes and levels of emotional closeness \cite{granovetter1973strength, onnela2007analysis}. Personal social networks typically consist of a handful of close and emotionally intense relationships (strong ties), essential for our well-being and health  \cite{house1988social, holt2010social, wittig2008focused, Litwin2014}, and a larger number of less close relationships (weak ties) that provide us with diversity and opportunities. Weak ties are crucial for the integrity of society since they function as bridges between otherwise separate parts of the social web \cite{granovetter1973strength, korte1970acquaintance, csermely2006weak, simmel1955conflict}.

Tie strength diversity can be quantified by analyzing the features of egocentric networks \cite{saramaki2014persistence, heydari2018multichannel, miritello2013time}. An egocentric network (ego-net) consists of an individual (the ego) and the social ties connecting the ego to its direct neighbors (alters)\cite{Marsden1990}. Such networks can be readily constructed from communication data, where the frequency of interaction is typically used as a proxy of tie strength~\cite{onnela2007structure}. Several studies have provided evidence of tie strength heterogeneity in ego-nets inferred from mobile phone data \cite{saramaki2014persistence, heydari2018multichannel, li2022evidence}, with recent findings on multiple online platforms indicating strong similarities in ego-net structural patterns across communication channels \cite{iniguez2023universal}. 

The extent of tie strength heterogeneity in ego-nets varies from person to person. Whatever the communication channel, there are distinct individual differences that are similarly distributed across channels~\cite{iniguez2023universal}. These individual patterns persist over time\cite{saramaki2014persistence, heydari2018multichannel,li2022evidence}, even if there are major changes in the composition of alters. In other words, people have \textit{social signatures}~\cite{saramaki2014persistence} and their networks tend to retain their characteristic feature of heterogeneity, independently of who their alters are.

One candidate explanation for this surprising universality -- of both tie strength heterogeneity and of the distribution of individual variation across channels -- is that it reflects some ego-level latent variable, whose distribution in the population determines the variety of ego-net tie strengths. Here, personality traits\cite{centellegher2017personality} are an obvious candidate to investigate.
However, people communicate on different channels with varying levels of activity\cite{PhysRevE.94.052319}, resulting in variations in network size that might obscure the role of personality traits in ego-net evolution.

Thus, we need a way of comparing personal networks that captures the similarity of tie strength heterogeneity despite differences in network size. This would help us clarify whether the methods used in earlier work (e.g.,\cite{saramaki2014persistence}) 
really capture the persistence of tie strength heterogeneity, rather than the persistence of network size alone. To this end, we explore four methods of measuring differences in heterogeneity between pairs of ego-nets and examine their sensitivity to ego-net size (i.e., the network degrees of the egos).

We demonstrate that some of the studied measures are less sensitive to ego-net degree and are therefore a better option for comparing the levels of tie strength heterogeneity of networks of different sizes. We also apply two of the measures to multichannel data on mobile telephone call and text message (SMS) logs, finding that the heterogeneity levels of egos are moderately correlated across these two channels. This observation supports the notion of an individual-level latent variable shaping ego-net characteristics and evolution.

\begin{figure}[t]
\includegraphics[scale=0.5]{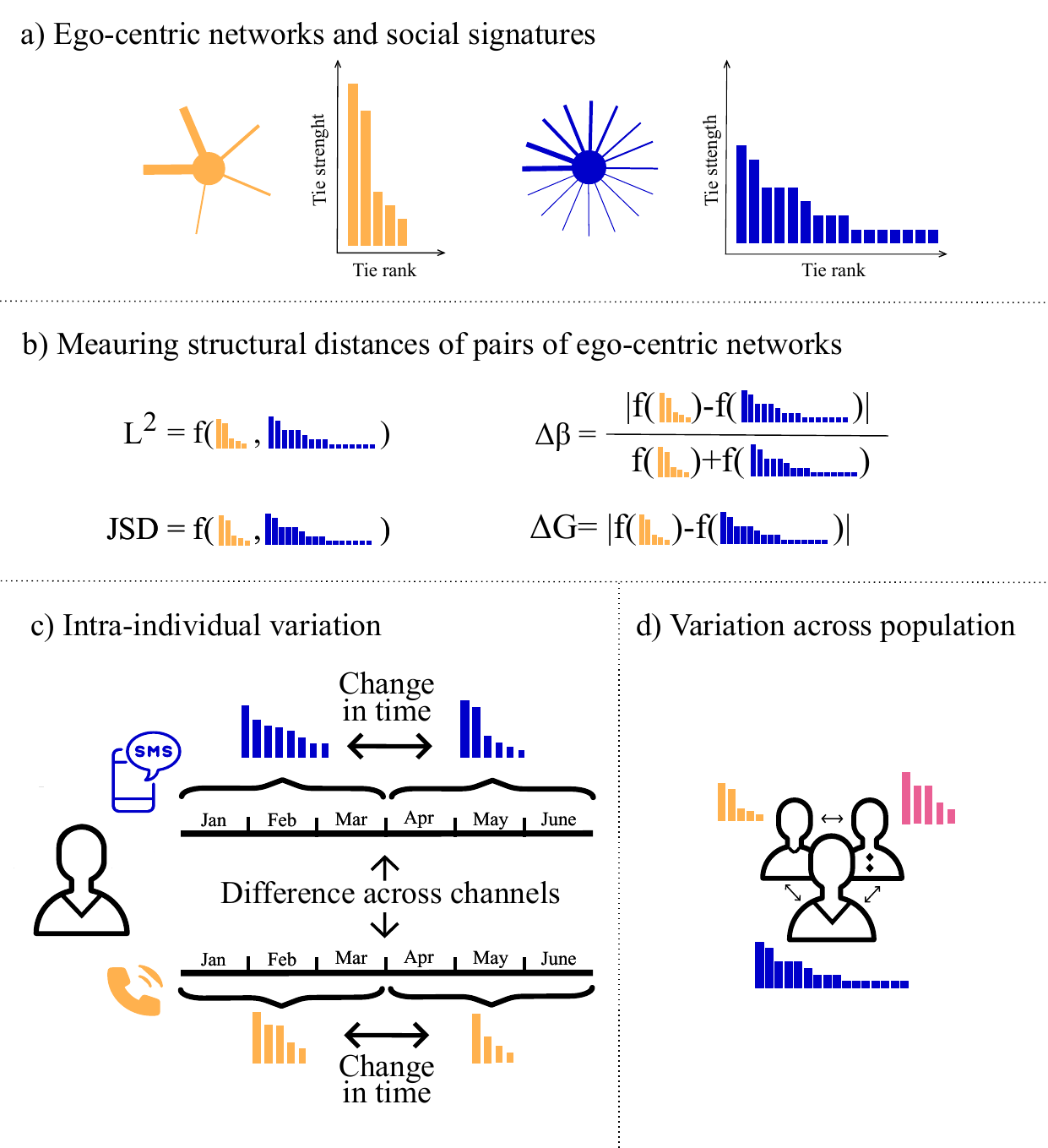}
\centering
\small
\caption{{\bf Ego-nets and distance functions for measuring their structural differences.} {\bf(a)} Two weighted ego-nets and their corresponding non-normalized social signatures. We construct an ego's social signature by sorting ties in descending order of strength. {\bf(b)}  Distance functions for measuring structural differences between ego-net pairs. The \(L^2\) and JSD distances directly assign a value to each ego-net pair. In contrast, when measuring the distance using the other two functions, preferentiality $\beta$ and the Gini coefficient $G$, a heterogeneity value is first calculated for each ego-net, and then the distances are measured. {\bf (c)} Intra-individual variations (self-distances) of an ego are calculated by measuring the distances between its ego-nets in two consecutive windows (change in time) and by measuring the distance between its ego-nets across communication channels (difference across channels). {\bf(d)} To determine whether the self-distances are smaller or larger than expected, we need to compare them with a reference. We make the reference distance distribution by calculating the distance between ego-nets of random pairs of egos in the population.
}
\label{fig:schematic}
\end{figure}

\section{Data and Methods}

\subsection{Data}
\label{subsec:data}
We use two mobile phone datasets in this study, both including call and short message (SMS) logs. The larger dataset includes communication logs of more than five million users over a six-month period and dates back to 2007, a time when mobile phone calls and SMS were more widely used, as smartphone-app-based alternatives did not yet exist. We refer to the larger data as the Large Mobile Phone (LMP) dataset and use it to examine the sensitivity of ego-nets' structural similarity measures to ego-nets' size difference. The smaller dataset is known as the Copenhagen Networks Study (CNS) dataset~\cite{sapiezynski2019interaction} and consists of communication logs between a group of around one thousand university students in 2012-2013.  After determining the robustness of the measures for ego-net comparison, we use both datasets to compare heterogeneity across different communication channels. 
For a detailed description of these two datasets, see Supplementary Information (SI) \srefsi{1.1} and \srefsi{1.2}.

\subsection{Egocentric networks and social signatures}
An egocentric network consists of the ego and the ego's alters, i.e., friends and family members that have direct social ties with the ego. In weighted ego-nets, the links are associated with weights that represent tie strength, thus being proxies of relationship intensity. There are several approaches for defining the weights using communication logs  \cite{urena2020estimating, heydari2018multichannel}. In this paper, we use the accumulated frequencies of communication events (numbers of calls or messages) during a certain time window as the link weights, similarly to approaches used in \cite{onnela2007structure, saramaki2014persistence, miritello2013time, wang2011human}. See \fref{fig:schematic} for schematic visualizations of two example ego-nets.

People vary in how they distribute their communication resources among their alters. To capture this inhomogeneity, the notion of \textit{social signatures} was introduced\cite{saramaki2014persistence} as the fraction of communication dedicated to alters when they are ranked in decreasing order of communication frequency. For the weighted ego-net of ego $e$, where $w_{ei}$ denotes the link weight of the $i$th most contacted alter, the vector representing the social signature is defined as
\begin{equation}
s_e=(w_{e1}/\sum_{i=1}^{k}{w_{ei}},\ldots,w_{ek}/\sum_{i=1}^{k}{w_{ei}}),\label{eq:signature}
\end{equation}
where $k$ is the degree (total number of alters) of the ego $e$.
Social signatures have been shown to be persistent over time, implying that people's networking patterns have distinct signatures\cite{saramaki2014persistence}. Moreover, people's call and SMS signatures have been shown to resemble each other, providing further evidence for the existence of distinct individual social signatures~\cite{heydari2018multichannel}.

\subsection{Measures of ego-net structural distance}
\label{subsection:heterogeneity_measures}

People vary in their social network size, in how actively they use a channel for communication, and in how they distribute their communication resources among their alters. Here, we will review four different functions that have been used in the literature to measure the structural distances between ego-nets. These functions take tie strength heterogeneity into account in different ways. In the subsequent Sections (\ref{subsection:persistence_degree_dependency} and \ref{subsection:self-similarity}), we will investigate whether the observed phenomena of ego-net persistence and across-channel similarity (reported in \cite{saramaki2014persistence, heydari2018multichannel, li2022evidence, iniguez2023universal}) hold true, irrespective of the extent to which the degrees of the ego-nets change over time (or differ across channels) as well as the distance function used for measuring the similarity.

\noindent \emph{\(L^2\) distance}\\
To measure the \(L^2\) distance (or Euclidean distance) between two social signatures, we first zero-pad the shorter signature, appending zeroes to its end in order to make sure that both signatures are of the same length. Then, the \(L^2\) distance between social signatures $s_1$ and $s_2$ is defined as
\begin{equation}
L^2(s_1,s_2) = \sqrt[]{\sum_{r=1}^{k} |f_{1r}-f_{2r}|^2},
\end{equation}
where $f_{1r}$ is the fraction of communication that the alter of rank $r$ in signature $s_1$ receives.

\noindent \emph{Jensen-Shannon distance}\\
To measure the Jensen-Shannon distance between two social signatures, we again first zero-pad the shorter one so that both of the signatures are of the same length. Then, the JSD distance $d_\mathrm{JSD}$ between social signatures $s_1$ and $s_2$ is defined as
\begin{equation}
d_\mathrm{JSD}(s_1, s_2) = {\big[H(\frac{1}{2}s_1 + \frac{1}{2}s_2) - \frac{1}{2}(H(s_1)+H(s_2))\big]}^{\frac{1}{2}},
\end{equation}
where $H(s_1)$, the Shannon entropy of $s_1$, is
\begin{equation}
H(s_1) = -\sum_{r=1}^{k}f_{1r}\ln{f_{1r}},
\end{equation}
with $k$ the maximum rank and $f_{1r}$ the fraction of communication dedicated to alter of rank $r$. ($f\ln f$ is taken as $0$ for $f=0$.)

\noindent \emph{Gini Coefficient distance}\\
The Gini coefficient~\cite{gini1921measurement} is commonly used to measure the deviation of wealth distribution from perfect equality, with values ranging from 0 (indicating complete equality) to 1 (indicating complete inequality). In our study, we utilize the Gini coefficient to quantify the inequality in the distribution of communication among alters of an ego, similarly to \cite{bhattacharya2016sex}.

If the link weights of an ego-net with degree $k$ are sorted in ascending order ($w_1 < w_2 < ... < w_k$), then the Gini coefficient $G$ is calculated as
\begin{equation}
\label{eq:gini_coef}
G = \frac{2\sum_{i=1}^{k}iw_{i}}{k\sum_{i=1}^{k}w_i} - \frac{k+1}{k}.
\end{equation}
We define the metric distance between a pair of Gini coefficients $G_1$ and $G_2$ as
\begin{equation}
\label{eq:GiniDistance}
\Delta G(G_1, G_2) = |G_1 - G_2|.
\end{equation}

\noindent \emph{Alter-preferentiality distance}\\
The alter-preferentiality parameter $\beta$ ~\cite{iniguez2023universal} is a measure defined for an ego-net growth model that incorporates a tuning parameter that determines the tendency of an ego to recurrently contact previously contacted alters, resulting in a ``rich-gets-richer'' phenomenon and leading to a heterogeneous ego-net, as opposed to choosing an alter uniformly at random resulting in a homogeneous ego-net. The range of $\beta$ is from 0 to infinity, where $\beta < 1$ corresponds to the homogeneous regime and $\beta > 1$ corresponds to the heterogeneous regime.

To estimate the alter preferentiality $\beta$ associated with a weighted ego-net, we utilize a maximum likelihood estimation method and a goodness-of-fit (GOF) test, as described in Ref.~\cite{iniguez2023universal}. Our GOF test uses Kolmogorov-Smirnov statistic~\cite{press1988numerical} with a p-value threshold of 0.1. Not all the fitted values pass the test, and therefore, the preferentiality value cannot be assigned to all the egos. The percentages of egos with valid fits are presented in Table \ref{table:data_size}. After fitting the $\beta$ values, we filter out the outliers by excluding the largest five percentile of $\beta$ values for all the analyses with the exception of Figure~\ref{fig:across_channel_correlation} for which because of the small size of the CNS dataset we skip this step. 

Unlike the Gini coefficient, which has a bounded range, $\beta$ has no upper bound. Therefore, we propose a normalized distance function for measuring the distance between a pair of alter-preferentiality parameters, $\beta_1$ and $\beta_2$, as

\begin{equation}
\label{eq:NormBetaDistance}
\Delta \beta(\beta_1, \beta_2) = \frac{|\beta_1 - \beta_2|}{\beta_1 + \beta_2}.
\end{equation}
Similarly to \eref{eq:GiniDistance}, this normalized distance is bounded between zero and one.

\subsection{Persistence, revisited}
\label{subsection:persistence}
Ego-nets are dynamic: people initiate new social relationships and lose contact with old acquaintances over time\cite{saramaki2014persistence, roy2022turnover}. Moreover, social ties usually display bursty patterns of activity: there are bursts of frequent contact that are separated by longer periods of silence \cite{oliveira2005human, candia2008uncovering, barabasi2005origin, karsai2011small}. Despite the constant changes and turnover in the sets of alters, several studies have reported structural stability in the shape of ego-nets \cite{saramaki2014persistence, heydari2018multichannel, li2022evidence}. There are two main aspects to the shape of ego-nets: the size or degree (the number of alters) and the manner in which the limited communication resources are distributed among the alters. However, it is not entirely clear whether the ego-net persistence reported in the literature is merely due to stability in ego-net size or whether the ego's pattern of tie strength heterogeneity is also persistent.

To answer this question, we must first define and quantify the persistence of ego-net structure. Persistence is commonly measured by comparing the temporal changes in the structures of individual ego-nets with the variation in population (e.g. the distribution of inter-individual distances) to determine if the changes are small or large (a method used in \cite{saramaki2014persistence, heydari2018multichannel, centellegher2017personality, li2022evidence, koltsova2021social, iniguez2023universal}). In this study, we take a similar approach but with some modifications which enable us to disentangle the effects of degree stability and tie-heterogeneity persistence.

To measure the temporal change for each ego and communication channel, we divide the timeline into two consecutive windows. In this study, instead of having consecutive windows with equal duration, we divided the timeline so there are equal numbers of communication events in each window. Then, similarly to the previous studies, we construct the weighted ego-net associated with each window and measure the structural difference between the two ego-nets, $d(ew_1, ew_2)$, using the distance functions of our choice (distance functions are listed in Section \ref{subsection:heterogeneity_measures}).

To investigate the effect of the degree change, we calculate the persistence relative to two different reference distributions: a general reference with no constraint (which is the approach taken in the earlier studies~\cite{saramaki2014persistence, heydari2018multichannel, centellegher2017personality, li2022evidence, koltsova2021social, iniguez2023universal}) and a degree-stratified reference consisting exclusively of the ego pairs that have a degree difference equal to the degree change of the ego for whom we are calculating the persistence. In the first approach, the reference distribution comprises the ego-net distances between randomly selected egos $x$ in the first period and randomly selected egos $y$ in the second period, without paying attention to their degrees (with $x \neq y$). In the second method, we form the reference distribution of a subset of these instances, namely the pairs $(x, y)$ that satisfy the condition $|k_{e1} - k_{e2}| = |k_{x1} - k_{y2}|$, where $k_{e1}$ and $k_{e2}$ are the degrees of ego $e$ in the different time windows, and $k_{x1}$ and $k_{y2}$ are the degrees of egos $x$ and $y$ in their respective time windows.

To quantify the structural persistence of the network of ego $e$, we calculate the $z$-score, a signed dimensionless quantity that measures the deviation of $d(ew_1, ew_2)$ from the mean value of the reference distribution of choice, $\mathrm{ref}$, in units of the standard deviation of $\mathrm{ref}$:
 
 \begin{equation}
 \label{eq:persistence_same_ref}
z(d(ew_1, ew2), \mathrm{ref}) = \frac{d(ew_1, ew2) - \mathrm{avg}(\mathrm{ref})}{\mathrm{std}(\mathrm{ref})}.
\end{equation}

A negative value means that the temporal change for the ego is less than the average distance between a random ego in the first window and another random ego in the second window, and this non-randomness is taken as indicative of ego-net persistence.

\section{Results}

Earlier studies have produced some evidence of ego-net persistence and of their similarities across communication channels, as indicated by intra-individual distances being smaller than the average distances between random individuals \cite{saramaki2014persistence, heydari2018multichannel, centellegher2017personality, li2022evidence, koltsova2021social, iniguez2023universal}. Here, we ask if the observed persistence is merely caused by the stability of ego-net size or if egos are also persistent in the manner in which they distribute their limited communication resources among their alters. We use the LMP dataset (see \ref{section:large_data}) which includes communication logs of over four million users for our investigations. This size facilitates computing degree-stratified reference distributions, as pairs of egos with a chosen degree difference can almost always be found.

First, in Section \ref{sec:Degree_difference_and_its_impacts}, we establish our motivation for probing the notion of ego-net stability by showing that all different distance functions introduced in Section \ref{subsection:heterogeneity_measures} are to some extent sensitive to the degree differences of the ego-net pairs. We also show that as compared with the variation in the population, people tend to have stable degrees in time and, to a smaller extent, similar degrees across the call and SMS channels.

Having established the motivation for revisiting the persistence of ego-nets, in Section \ref{subsection:persistence_degree_dependency} we show that the structural persistence reported in the literature holds when measured as compared to references with similar degree differences and is not a mere side effect of degree stability.

Moreover, we observe that among the four structural distance functions used in the literature (see Section  \ref{subsection:heterogeneity_measures}), the distance functions based on the Gini coefficient and the alter preferentiality $\beta$ are less sensitive to network size differences. Thus, they are more appropriate measures for comparing ego-nets of different sizes if we wish to focus on tie strength heterogeneity.

Finally, armed with measures relatively insensitive to ego-net size, we investigate the similarity of networks of egos across call and SMS communication channels. For this purpose, we use the CNS and LMP datasets (see Sections \ref{section:copenhagen_data} and \ref{section:large_data} for data description). The comparison over these channels suggests that the heterogeneity levels of ego-nets are moderately correlated across call and SMS channels.

\subsection{Why do we need to disentangle the persistence of ego-net size and tie strength heterogeneity?}
\label{sec:Degree_difference_and_its_impacts}

We are interested in measuring the persistence of tie strength heterogeneity of the ego-nets because we hypothesise that the distribution of tie strength is rather stable and possibly determined to a large extent by individual personality traits. To prove this, we need to compare ego-nets of the same individual across time and different channels. However, measures applied to quantify the difference between ego-nets may be sensitive not only to how tie strength is distributed in the ego-nets but also to their degrees. Ego-nets of similar degrees may appear similar purely because of the small degree difference. We wish to minimise this effect and we do so by investigating the sensitivity of the distance functions on degree variation in order to choose the less sensitive measures.

We empirically investigate how the distance measures (introduced in Section \ref{subsection:heterogeneity_measures}) depend on the degree difference of the ego-net pairs that are being compared. To this end, we take a random sample of approximately 7,000 egos from the LMP dataset and form all possible ordered ego pairs ($\sim$ 7000*6999 pairs). For each ego pair $(e_1, e_2)$, we construct an ego-net for $e_1$ based on the first half of its call log and an ego-net for $e_2$ using the second half of $e_2$'s call log. Then we calculate the distance between these pairs of ego-nets using each of the four distance functions defined in Section \ref{subsection:heterogeneity_measures} (notice that these are the same ego-nets that we will use to form the reference distribution in Section \ref{subsection:persistence_degree_dependency}).

\begin{figure}[htbp]
\small
\includegraphics[scale = 0.5]{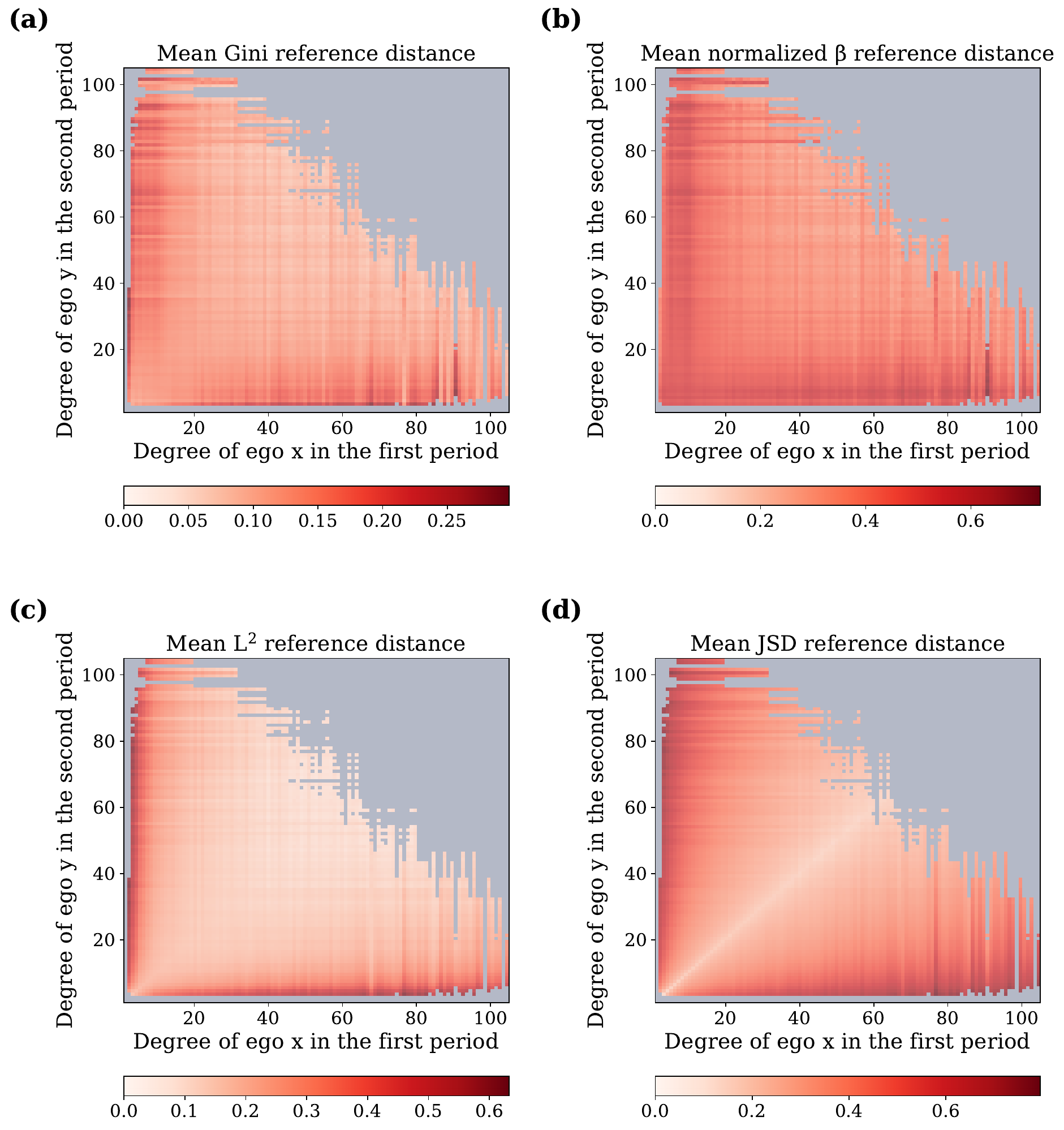}
\centering
\caption{{\bf Average measured distance is smaller for pairs with similar degrees.} The distribution of average reference distances (distance between the network of a randomly selected ego, $x$, in the first period and the network of another randomly selected ego, $y$, in the second period) measured by different distance functions [{\bf (a)} Gini distance, {\bf (b)} normalised $\beta$ distance, {\bf (c)} $L^2$ distance, and {\bf (d)} Jensen-Shannon divergence distance]. For all the distance functions, the average reference distance tends to increase as the degree difference becomes larger. However, the effect is sharper for the JSD and $L^2$ distance functions.}
\label{fig:ref_distance_heatmap}
\end{figure}

To investigate how the measured distance depends on the degrees, we group the ego-net pairs based on their degrees ($k_1$, $k_2$) and calculate the average distance for each group and each distance function. Fig. \ref{fig:ref_distance_heatmap} shows the average distances for each group as a function of the degree of the ego-nets in four heatmaps, one for each distance function. Bins with too few pairs (less than 200 pairs) are not shown. The plots reveal that for all distance functions, the measured distance tends to be smaller when the ego-nets have similar degrees. However, the effect is stronger for the two signature-based distance functions (JSD and $L^2$ distances).

\begin{figure}
\includegraphics[scale=0.4]{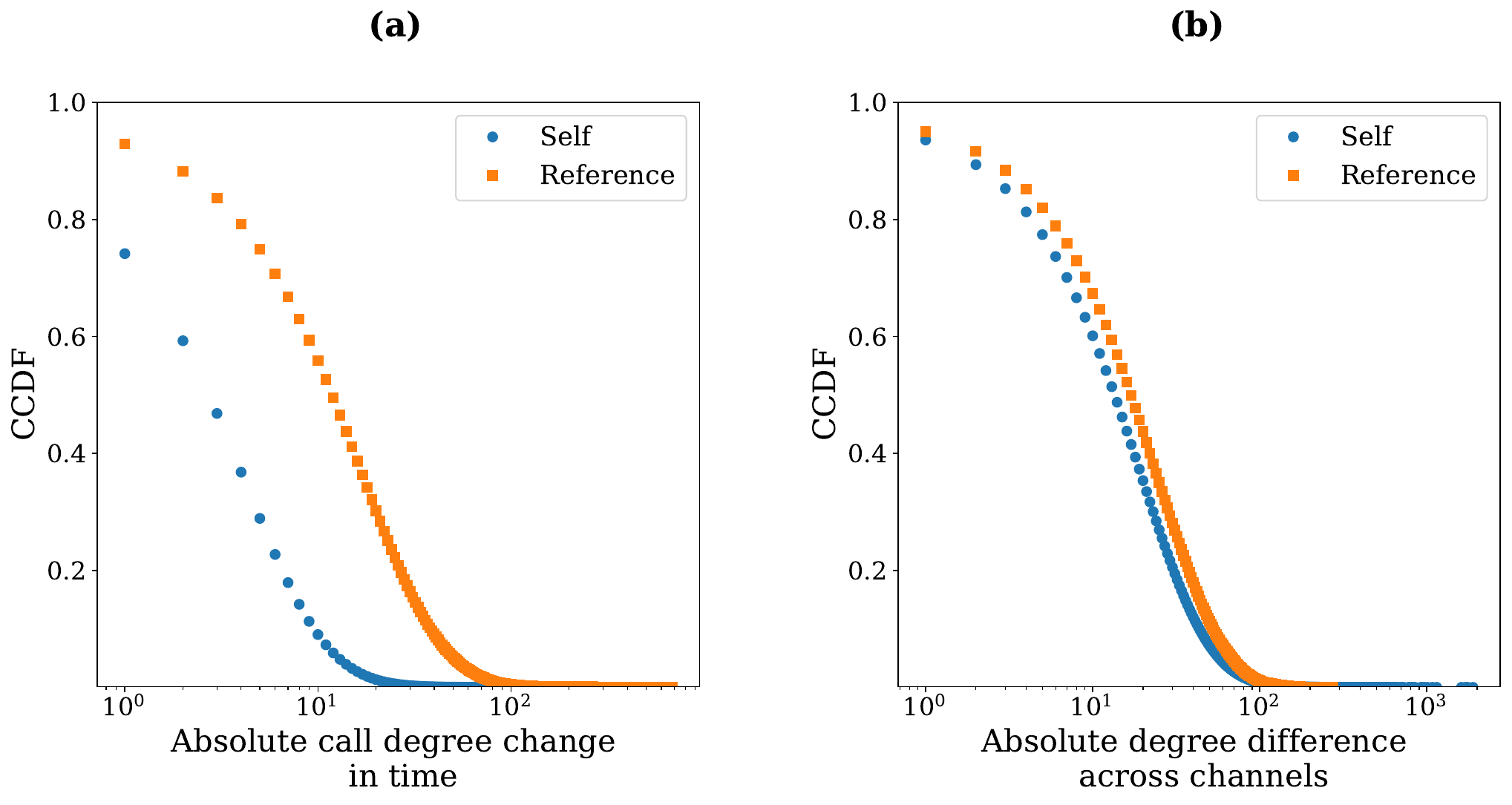}
\centering
\caption{ {\bf Egos have stable degrees in time and to a smaller extent across call and text channels.} Left plot: The complementary cumulative distribution (CCDF) of degree change over time of call ego-nets in the large mobile phone dataset (blue circles) vs. the CCDF of degree difference of the random ego-net pairs (orange squares). The comparison of the two curves shows that egos have persistent call degrees as compared to the call degree variation in the population. {\bf (right)} CCDF of the personal call and SMS degree difference of egos in the LMP dataset (blue circles) vs. the degree difference of the call and SMS ego-nets belonging to random pairs of egos (orange squares). The comparison shows that egos tend to have smaller differences between their call and SMS degrees as compared to the random pairs. However, the effect is weaker than the call degree persistence case.}
\label{fig:self_vs_ref_delta_k}
\end{figure}

As the structural distance of ego-nets and degree difference appear correlated, the next natural step is to study the individual and population variations of degree.
\fref{fig:self_vs_ref_delta_k} shows complementary cumulative distributions of both the amount of individual degree change of ego-nets in time and the individual degree differences of ego-nets across channels, as compared to the degree differences of random ego-net pairs (the population variation). We observe that compared to the population variation, egos tend to have stable degrees in time and, to a smaller extent, similar degrees across the call and SMS channels.

The degree sensitivity of the distance functions, along with the degree stability of the egos, prompt us to revisit the notions of ego-net persistence and similarity and quantify how degree differences affect ego-net persistence as measured with different distance functions (see Sections \ref{subsection:persistence_degree_dependency} and \ref{subsection:self-similarity}).

\subsection{Impact of degree change on the structural persistence of ego-nets}
\label{subsection:persistence_degree_dependency}

Next, we wish to investigate the impact of degree changes on the persistence of ego-nets when using different functions to measure structural distance (see Section \ref{subsection:heterogeneity_measures} for a description of different distance functions and Section \ref{subsection:persistence} on how to quantify persistence). For this analysis, we utilize the call log data from the LMP dataset. The extensive size of the dataset enables us to draw a sufficiently large sample for establishing a representative reference distribution of distances in the population. To quantify persistence, we compare the self-distances that represent the structural change of each ego-net from one period to another against the reference distribution and calculate the corresponding $z$-scores [see \eref{eq:persistence_same_ref}].

When calculating the self distances, we limit ourselves to the egos to whom we can assign both $\beta$ and Gini heterogeneity values, i.e. those ego-nets that pass the goodness-of-fit test for $\beta$ and have degrees larger than one so it is possible to calculate their Gini coefficient. This leaves us with around 3 million ego-nets for whom we calculate the structural distance between their ego-nets in two consecutive time windows, according to different measures.

To determine if the self-distances are smaller or larger than expected, as explained in Section \ref{subsection:persistence}, we use a general, degree-agnostic reference and $|\Delta k|$-specific references. The sample for making the degree-agnostic reference is the same as in \ref{sec:Degree_difference_and_its_impacts} ($\sim 49$M pairs).
As for $|\Delta k|$-specific references, we use a specific subset of the sample according to the degree change of the ego whose ego-net persistence we are assessing (See \ref{subsection:persistence}).

\begin{figure}[htbp]
\small
\includegraphics[scale = 0.34]{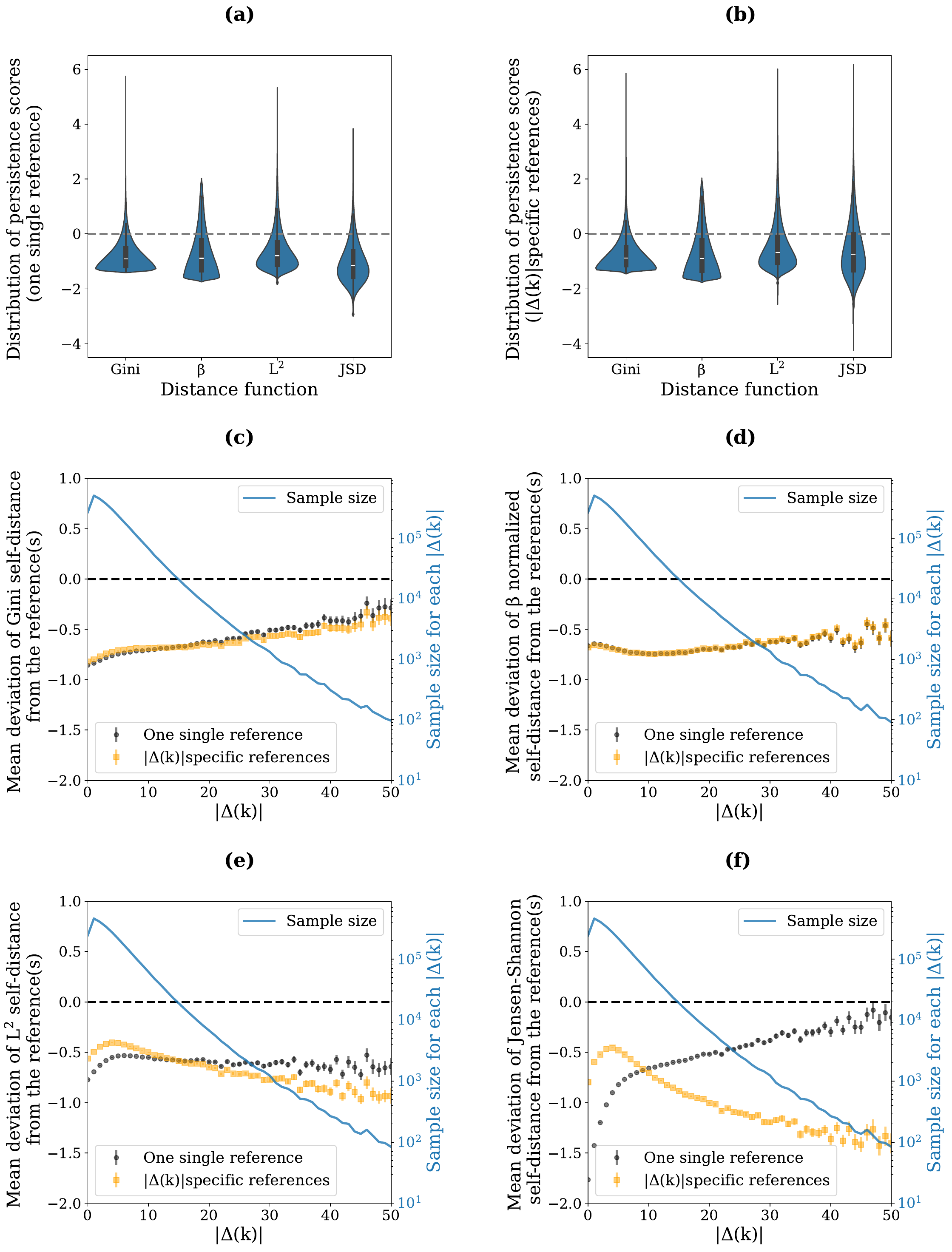}
\centering
\caption{{\bf Structural persistence of call ego-nets in the LMP dataset using different functions for measuring structural distance.} {\bf (a)} Distribution of persistence scores in the LMP dataset when measured against one single reference and in {\bf (b)} when measured against $|\Delta(k)|$-specific references. In both cases, most of the scores are negative, which indicates the stability of call ego-nets.
Average persistence as a function of $|\Delta(k)|$ (absolute degree change) is shown in {\bf (c)} using the Gini distance function; in {\bf (d)} using the normalised $\beta$ distance; in {\bf (e)} using the $L^2$ distance; and in {\bf (f)} using the JSD distance. 
The average persistence scores (the deviation of self-distances from reference distances) are visualised in grey if they are measured against one single reference distribution (regardless of $|\Delta(k)|$) and in orange if against $|\Delta(k)|$-specific reference distributions. See Eq. \ref{eq:persistence_same_ref} for the definition of the persistence score. All four heterogeneity measures have negative average values across the degree range. Moreover, the persistence plots for the Gini coefficient and $\beta$ do not change much when switching from a single reference to $|\Delta(k)|$-specific references. This suggests that Gini and $\beta$ are better measures for comparing ego-nets of different sizes. The blue curves in plots (a), (b), (c), and (d) show the number of egos with a particular absolute degree change ($|\Delta(k)|$) from the first period to the second period (values are on the right axes).}
\label{fig:persistence}
\end{figure}

Having the self-distances and the reference distributions, using Eq.~(\ref{eq:persistence_same_ref}), we can measure the persistence score of each ego. Fig.~\ref{fig:persistence} shows the distribution of persistence scores of call ego-nets among the population (around three million egos) in the LMP dataset. The persistence values are calculated using all four distance functions introduced in Section \ref{subsection:heterogeneity_measures} and both against a single degree-agnostic reference [panel (a)] as well as $|\Delta k|$-specific references [panel (b)]. We observe that for all these cases, most egos have negative persistence scores, indicating the stability of the ego-nets.

So far, we have quantified the level of persistence whose existence was statistically observed in the previous studies (\cite{saramaki2014persistence, heydari2018multichannel, centellegher2017personality,li2022evidence}) and also shown that the reported persistence holds when controlling for different levels of degree change and thus it is not merely a side effect of degree stability. To examine the impact of degree change on the persistence of the ego-nets, we take a closer look at the persistence scores by grouping the egos based on the absolute value of their degree change, denoted as $|\Delta(k)|$, and calculate the average persistence for each group separately (see \fref{fig:persistence}). We observe that the average of the persistence scores measured against both a single reference (the grey scatter plots) and $|\Delta k|$-specific references (the orange scatter plots) remains relatively constant as a function of $|\Delta(k)|$ for the normalized $\beta$ distance and $L^2$ distances, whereas a clear dependence on $|\Delta(k)|$ can be observed for the JSD distance.

Moreover, we observe that when switching from one single reference to $\Delta k$-specific references, the persistence scores for the $L^2$ and JSD distances change more significantly as compared to the $\beta$ and Gini persistence scores (see the orange scatterplots in \fref{fig:persistence} in comparison to the grey scatterplots). This result indicates that the alter preferentiality $\beta$ and the inequality measure $G$ are both fairly insensitive to degree difference and, thus, can be considered more suitable for comparing ego-nets with varying sizes.

\subsection{Application: similarity of ego-nets across call and SMS channels using the Gini coefficient and the alter preferentiality $\beta$}
\label{subsection:application_on_CNS}

Finally, we apply our understanding of the proper measures for ego-net comparison to analyse the cross-channel similarity of ego-nets. This is motivated by the fact that an ego-net reconstructed from data on communication on one single channel only represents an incomplete reflection of the underlying social network. Then, the accuracy and completeness of such reflections are determined by how exclusively and actively the ego uses the communication channel and what portion of observed events are related to actual social relationships (e.g., calling a service line or calling a wrong number are not).

We start by using the LMP dataset to assess the similarity of ego-nets across channels with an approach that controls for different levels of degree difference (similar to our approach in analyzing the persistence of ego-nets in Section~\ref{subsection:persistence_degree_dependency}). We conclude that call and SMS ego-nets resemble each other, but the effect is smaller than the observed persistence of ego-nets in time (see Section \ref{subsection:self-similarity} for details). 

\begin{figure}
\includegraphics[scale = 0.5]{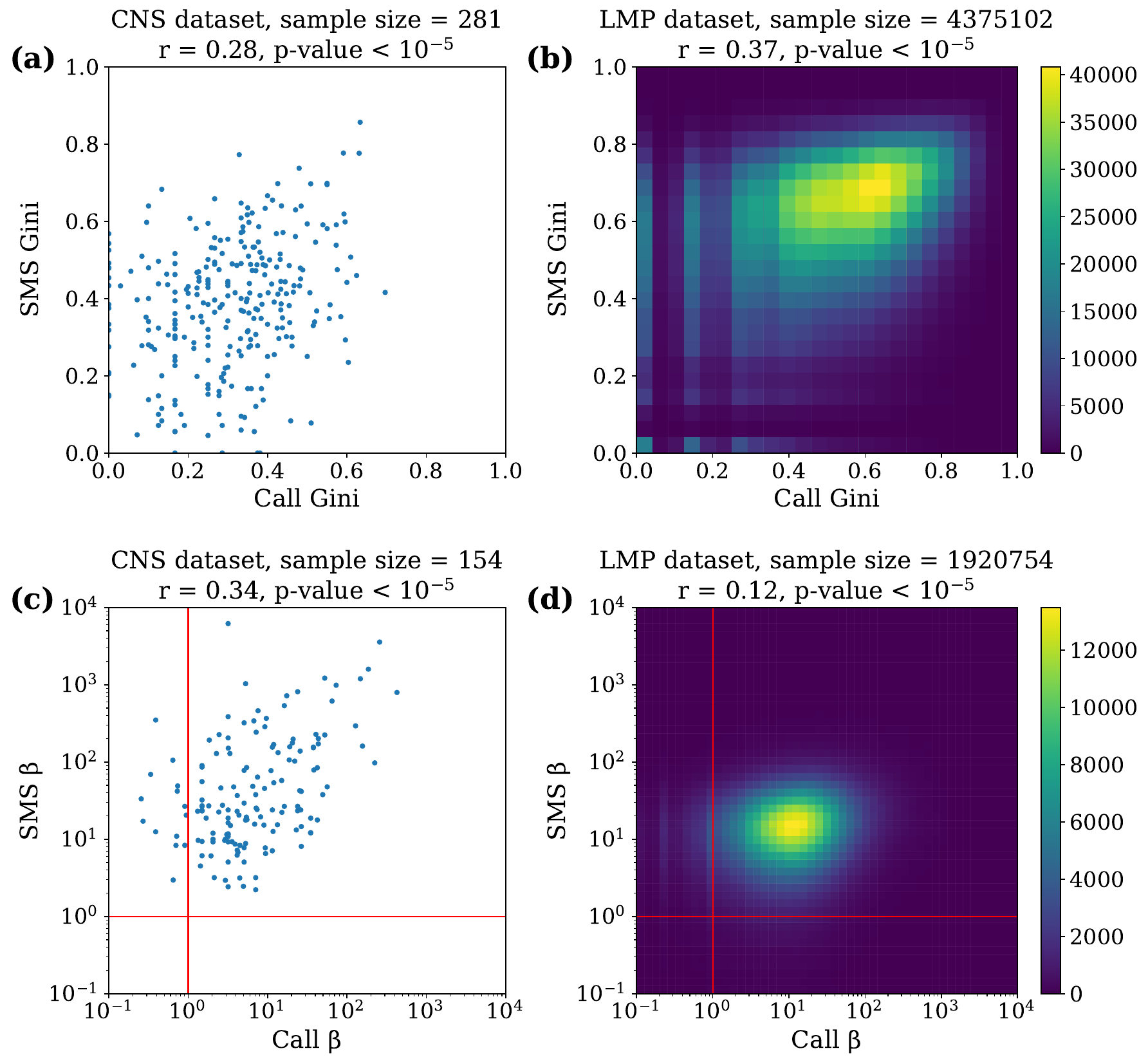}
\centering
\caption{{\bf Call and SMS ego-nets are correlated in their heterogeneity levels.} Gini values associated with tie-strength heterogeneity of SMS ego-nets of the same individuals are plotted as a function of their call Gini coefficient in {\bf (a)} the CNS dataset and in {\bf (b)} the LMP dataset (colours in the heatmap indicate the number of egos falling into each bin regarding their call and SMS Gini values). Panels {\bf (c)} and {\bf (d)} show similar plots for the alter parameter ($\beta$). The red lines indicate $\beta = 1$, the border between homogeneous and heterogeneous regimes. We observe that most of the egos fall into the heterogeneous regime. The positive Pearson correlation values in all four plots show that SMS and call heterogeneity values are correlated.}
\label{fig:across_channel_correlation}
\end{figure}

Next, as the Gini coefficient and the alter preferentiality $\beta$ do not require the use of reference distributions, we directly correlate the Gini and $\beta$ values for call and SMS based ego-nets, for both of our datasets (LMP and CNS), as shown in Fig.~\ref{fig:across_channel_correlation}. We observe moderate correlations across the modalities, indicating that there are clear similarities between an individual's tie strength heterogeneity across the channels. 

Taken together, the above results support the hypothesis of a latent variable, such as a personality trait, giving rise to similar ego-net shapes for an individual across different channels. However, as the correlation is only moderate, this latent variable might not be the only determinant of tie strength heterogeneity in an ego-net.

\subsection{Discussion}
Egocentric networks are the building blocks of societal-level networks. They comprise a wide variety of relationships ranging from romantic partners and family to friends, colleagues, and acquaintances. This diversity of relationship types is reflected in the structure of ego-nets: not all social ties are equally strong, but tie strength is typically heterogeneous with a few strong and a larger number of weaker ties. This characteristic pattern has been observed in social networks inferred from various communication datasets \cite{saramaki2014persistence, miritello2013time, heydari2018multichannel, iniguez2023universal}. Within this broad pattern, there is, however, individual-level variation, as people express different levels of heterogeneity in their personal networks. Besides tie strength heterogeneity, ego-nets vary in size, i.e., network degrees (see, e.g.,~\cite{onnela2007structure}). To some extent, the degree of an ego-net reflects the size of the ego's underlying social circle, but it is also affected by empirical issues such as sampling (the communication channel used, and the length of the observation window \cite{PhysRevE.94.052319, krings2012effects}).

The heterogeneity of tie strength appears to be a universal characteristic of ego-nets~\cite{iniguez2023universal}: its distribution is similar in ego-nets derived from multiple communication channels. However, as the datasets in Ref.~\cite{iniguez2023universal} do not comprise the same individuals, it is not known what causes this similarity. On one hand, the similarity may be a population-level consequence of people having different channel preferences and behaving differently on different channels so that the mixture of these different behaviours is behind the distribution of heterogeneity. Alternatively, it may arise from 
latent-individual level variables, such as personality traits, which determine people's ego-net structure on any channel.

Both ego-net size and tie strengths are encapsulated in so-called social signatures~\cite{saramaki2014persistence} that quantify the share of communication that alters receive as a function of their rank in an ego-net. These signatures are rather persistent for an individual from one time period to another, even when the combination of alters of the ego changes\cite{saramaki2014persistence}. An individual's signatures also resemble each other across communication channels \cite{heydari2018multichannel}. However, until now, it has been unclear whether the persistence is due to degree stability alone or tie strength heterogeneity persists as well. In this paper, we carried out an analysis taking into account the sensitivity of distance functions used for comparing ego-nets on their degree differences. We showed that the observed persistence of social signatures and their individual-level similarity across the call and SMS channels are not merely a side effect due to the stability of the network size. While the effect is somewhat smaller, it is still statistically significant when controlling for degree differences. Our results indicate that although part of the persistence of the signatures of individuals indeed stems from the stability of the size of the ego-nets over time and channels, there is a considerable contribution to this persistence related to other individual features, possibly to personality traits.

Our comparison of the distance functions also revealed that two measures are far less sensitive to degree differences: the alter preferentiality $\beta$ and the Gini coefficient. These two are thus better options for assessing the persistence of tie strength heterogeneity when network sizes are also heterogeneous and for understanding whether the same individuals have similar ego-net shapes on different communication channels, where network sizes may generally vary quite extensively. These measures are highly correlated and theoretically related (see \ref{subsubsection:empirical_correlation}), and therefore, either of them can be used. The Gini coefficient is a well-known statistical indicator and is also straightforward to compute. The Gini coefficient can be calculated for any ego-net with a degree larger than one (which is also a prerequisite for any tie strength heterogeneity to be possible). On the other hand, the alter-preferentiality parameter $\beta$ has more explanatory power since it is associated with an ego-network growth model but is obtained via a rather cumbersome fitting procedure. Moreover, the fitting procedure automatically filters out ego-nets of too low degree and entirely homogeneous networks (see Fig. \ref{fig:fraction_egos_with_valid_beta_call}).

Another advantage of both coefficients is that they attribute a heterogeneity value to each ego-net, facilitating a more straightforward comparison than when using signature-based distance functions (e.g. JSD and $L^2$ distances). This is particularly relevant for comparing ego-nets across different communication modalities and for testing the latent-variable hypothesis. While some previous studies have examined the similarities and differences of ego-nets across different communication channels, the results have not been conclusive. Multichannel studies have shown that examining a single channel provides only a partial description of the ego's interactions, and a holistic approach is needed to gain a more realistic picture~\cite{PhysRevE.94.052319, zignani2015calling, heydari2018multichannel}. A study comparing call and SMS ego-nets pointed out that the composition and rankings of the alters in these two channels can be very different \cite{heydari2018multichannel}, even though the ego-nets of one individual are similarly shaped. Another study \cite{aledavood2016channel} compared temporal communication patterns of calling and texting and reported significant differences, which may indicate their different functionalities.

To this end, we applied the Gini and $\beta$ measures to investigate if the call and SMS ego-nets of an individual are correlated in terms of tie strength heterogeneity. We used two datasets that contain both calls and SMS messages and observed that the tie strength heterogeneity measures show a moderate correlation for call and SMS ego-nets, which is in line with the latent-variable/personality-trait hypothesis, where an individual tends to maintain similar networks on any media. However, the moderate level of correlation also leaves room for population-level explanations for the observed universality in heterogeneity distributions, as the hypothetical latent variable cannot be the sole determinant of tie strength heterogeneity in an ego-net. 
This points out that to properly understand the reasons for tie strength heterogeneity across different communication channels, there is a need for data involving the same subjects interacting on even more channels---one might also envision experimental setups where people repeatedly build and maintain networks on online channels and the tie strength heterogeneity in the resulting networks is investigated in detail.

\section{Declaration}
\subsection{Availability of data and materials}
The CNS data analysed in this paper is publicly available via {\it figshare} \cite{sapiezynski2019interaction, sapiezynski2019copenhagen}. While the LMP data is not publicly available and shared with us via a non-disclosure agreement, it has been extensively studied in the literature (see, for example, \cite{onnela2007analysis,onnela2007structure,karsai2011small,kivela2012multiscale,kovanen2013temporal,unicomb2018threshold,heydari2018multichannel}).
\subsection{Competing interests}
The authors declare that they have no competing interests.
\subsection{Funding}
This study was part of the NetResilience consortium funded by the Strategic Research Council at the Academy of Finland (grant numbers 345188 and 345183). G.I. and J.K. acknowledge support from AFOSR (Grant No. FA8655-20-1-7020), project EU H2020 Humane AI-net (Grant No. 952026), and CHIST-ERA project SAI (Grant No. FWF I 5205-N). J.K. acknowledges support from European Union’s Horizon 2020 research and innovation programme under grant agreement ERC No 810115 - DYNASNET.
\subsection{Author's Contributions}
S.H., G.I., J.K., and J.S. conceived, designed, and developed the study.  S.H. analysed the empirical data.  G.I. derived the analytical relations in Section \ref{subsubsection:theoritical_relation}.  S.H., G.I., J.K., and J.S. wrote the paper.
\subsection{Acknowledgments}
We acknowledge the computational resources provided by the Aalto Science--IT project.
\subsection{Abbreviations}
\textbf{Ego-net:} egocentric network,
\textbf{SMS:} a type of text message transmitted via Short Messaging Service,
\textbf{LMP:} Large Mobile Phone,
\textbf{CNS:} Copenhagen Networks Study,
\textbf{SI:} Supplementary Information,
\textbf{JSD:} Jensen-Shannon Divergence,
\textbf{GOF:} goodness-of-fit,
\textbf{CCDF:} Complementary cumulative distribution function.
\newpage



\appendix

\setcounter{figure}{0}
\setcounter{table}{0}
\renewcommand{\thefigure}{S\arabic{figure}}
\renewcommand{\thetable}{S\arabic{table}}
\renewcommand{\thesection}{S\arabic{section}}

\section{Communication data}
\label{sec:data}

\subsection{The Large Mobile Phone (LMP) dataset}
\label{section:large_data}
The Large Mobile Phone (LMP) dataset consists of time-stamped logs of outgoing communication of approximately 20\% of the population of an undisclosed European country. The user IDs are anonymised, and the logs span six months in 2007. The dataset was initially introduced in \cite{onnela2007analysis}. To ensure data quality, we have excluded self-communication events and records made by IDs listed under ``family contracts'' with the operator company, which could involve multiple users sharing the same phone line. This leaves us with more than 5 million users, 1.3 billion calls, and 613 million short messages (see Table \ref{table:data_size} for exact numbers). It is important to note that the data is directional, and we only use the outgoing communication events when constructing ego-nets, which is unlike our approach for the CNS dataset, where we also include incoming call/SMS to the ego-nets (see section \ref{section:copenhagen_data}). This is because the LMP dataset includes two types of individuals: company-users, for whom we have outgoing communication logs and construct the ego-nets, and non-company users, who only appear in the data if a company-user contacts them. Therefore, if all events are considered regardless of direction, the ties to non-company alters would systematically have lower weights as we lack information on outgoing communication events of those alters. While the data is not publicly available, it has been extensively studied in the literature (see, for example, \cite{onnela2007analysis,onnela2007structure,karsai2011small,kivela2012multiscale,kovanen2013temporal,unicomb2018threshold,heydari2018multichannel}).

\subsection{Copenhagen Networks Study (CNS) dataset}
\label{section:copenhagen_data}

The Copenhagen Networks Study (CNS)~\cite{stopczynski2014measuring,sapiezynski2019interaction} is a multichannel dataset collected via mobile phone devices distributed among around one thousand volunteer university students in 2012-2013. Data includes communication logs of calls and text messages (SMS) exchanged between the students as well as data on their physical proximity (captured by Bluetooth sensors)~\cite{stopczynski2014measuring}. Here, we use the portion of mobile phone communication data that is publicly available as described in~\cite{sapiezynski2019interaction}, e.g. one month's worth of call and SMS logs between anonymised users. Data is publicly available via {\it figshare} in~\cite{sapiezynski2019copenhagen}.

The call and SMS logs include only the events between the subjects in the study. We disregard the missed calls when making the ego-nets. Table \ref{table:data_size} shows the number of remaining egos and events in each channel. We define indirectional call and SMS ego-nets by adding a call or SMS between $x$ to $y$ to both $x$'s and $y$'s ego-net, regardless of who has initiated it. We define the tie strengths simply as the number of events of the corresponding type between the ego-alter pairs.

\begin{table}[!ht]
\small
\noindent\makebox[\textwidth]{ \begin{tabular}{l l r r r}
\toprule
Dataset & Communication channel & Number of egos & Number of events & \% of egos with valid $\beta$\\
\midrule
LMP & Phone calls & 5994967 & 1342862618 & 64\% \\
LMP & Short messages & 5387745 & 613751054 & 50\% \\
CNS & Phone calls  & 525 & 3234 & 37\% \\
CNS & Short messages (SMS)  & 568 & 24333 & 53\% \\
\bottomrule
\end{tabular}}
\caption{
\small A summary of the datasets used in this study. The large mobile phone (LMP) dataset is used in sections \ref{sec:Degree_difference_and_its_impacts}, \ref{subsection:persistence_degree_dependency}, \ref{subsection:application_on_CNS}, \ref{subsection:beta and Gini}, and \ref{subsection:self-similarity}, while the Copenhagen Network (CNS) study dataset is only used in section \ref{subsection:application_on_CNS}.
}
\label{table:data_size}
\end{table}

\section{Gini and $\beta$: The less sensitive heterogeneity measures}
\label{subsection:beta and Gini}

In Section \ref{subsection:persistence}, we observed that the Gini coefficient and the alter preferentiality $\beta$ are better choices for comparing ego-nets of different sizes. Here, we take a closer look at the two measures and compare them empirically and theoretically. We observe that these two measures are closely related both according to their mathematical definitions and also based on the empirical data.

\subsection{The empirical correlation}
\label{subsubsection:empirical_correlation}
\begin{figure}
\includegraphics[scale=0.4]{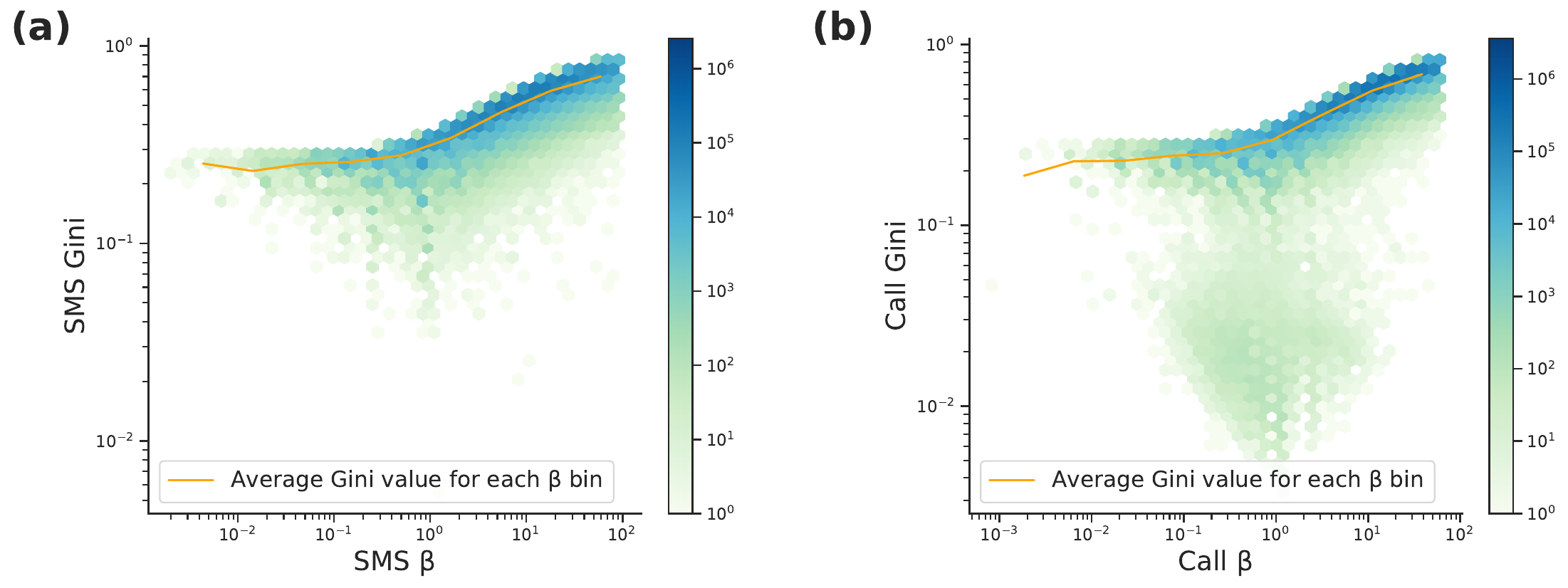}
\centering
\caption{{\bf Gini and $\beta$ are heavily correlated.} The colour intensity indicates the number of egos which fall into each bin regarding their $\beta$ and Gini coefficients. The ego-nets are made based on the call and SMS events in the LMP dataset (dataset described in \srefsi{1.2}). We observe that these two variables are highly correlated for both call and SMS ego-nets, with a Pearson correlation coefficient equal to 0.84 for the call ego-nets (sample size = 3626453 and $p < {10} ^ {-5} $), and 0.8 for SMS ego-nets (sample size = 2553079 and $p < {10} ^ {-5} $). We have excluded outlier $\beta$ values (the top five percentile).}
\label{fig:gini_and_beta_heatmap}
\end{figure}

To compare the heterogeneity values measured by the Gini coefficient and the alter preferentiality $\beta$, we check the Pearson correlation coefficient between the measured heterogeneity values of the same egos in the LMP dataset for both call and SMS channels. Correlation coefficients of $\sim 0.8$ show that these two values are strongly correlated, and the large sample size of a few million individuals and $p$-values smaller than $10^5$ show the statistical significance of the observed correlations (See \fref{fig:gini_and_beta_heatmap} demonstrating the correlation between these two measures for both call and SMS ego-nets in the LMP dataset).

\subsection{The theoretical relation}
\label{subsubsection:theoritical_relation}

In the minimal model of ego-net dynamics introduced in Ref. \cite{iniguez2023universal}, individuals allocate interactions via cumulative advantage and a tunable amount of random choice. At an initial time $\tau_0 = k a_0$ with $k$ the degree of the ego-net, all alters have minimal activity $a_0$. At any time $\tau \geq \tau_0$, the probability that an alter with activity $a$ becomes active at time $\tau+1$ (and thus $a \mapsto a+1$) is
\begin{equation}
\label{eq:connProbBeta}
\pi_a = \frac{a_r / t_r + \beta^{-1}}{ k ( 1 + \beta^{-1} ) },
\end{equation}
with $a_r = a - a_0$, $t_r = t - a_0$, and $t = \tau / k$ the mean alter activity. The alter preferentiality $\beta = t_r / \alpha_r$ (with $\alpha_r = \alpha + a_0$ and $\alpha$ a tunable parameter of the model) interpolates between two regimes: random alter choice ($\beta \to 0$ and $\pi_a \sim 1/k$), and preferential alter selection ($\beta \to \infty$ and $\pi_a \sim a_r / \tau_r$ with $\tau_r = \tau - \tau_0$). An analytical treatment of the model leads to the following expression for the activity distribution $p_a(t)$ (the probability that a randomly chosen alter has activity $a$),
\begin{equation}
\label{eq:actDist}
p_a(t) = p_0 \frac{ a_r^{-1} }{ \mathrm{B}( a_r, \alpha_r ) } \left( 1 + \frac{1}{\beta} \right)^{-a_r},
\end{equation}
where $p_0 = \left( 1 + \beta \right)^{-\alpha_r}$ and $\mathrm{B} (a_r, \alpha_r)$ is the Euler beta function \cite{iniguez2023universal}.

The Gini coefficient $G$ of \eref{eq:gini_coef} can be written in terms of the cumulative distribution function of alter activity, $P_a(t) = \sum_{a' = a_0}^a p_{a'}(t)$, by approximating $a$ as a continuous variable, writing the Lorenz curve of $P_a$ as an integral, and discretizing $a$ once more,
\begin{equation}
\label{eq:GiniActDist}
G = 1 - \frac{1}{t} \sum_{a = a_0}^{a_m} (1 - P_a)^2,
\end{equation}
where $p_a$ follows \eref{eq:actDist}. In the heterogeneous regime of preferential alter selection, the activity distribution approaches a gamma distribution with shape $\alpha_r$ and scale $\beta$ \cite{iniguez2023universal},
\begin{equation}
\label{eq:ActDistSmallAlpha}
p_a(t) = \frac{ 1 }{ \beta^{\alpha_r} \Gamma(\alpha_r) } a_r^{\alpha_r - 1} e^{ -a_r / \beta }, \quad \beta \to \infty,
\end{equation}
meaning that the Gini coefficient can be approximated as \cite{mcdonald1979analysis}
\begin{equation}
\label{eq:GiniApprox}
G = \frac{\Gamma (t_r/\beta + 1/2)}{\sqrt{\pi} \Gamma (t_r/\beta + 1)}, \quad \beta \to \infty,
\end{equation}
with $\Gamma$ the gamma function.

\begin{figure}
\includegraphics[scale=0.5]{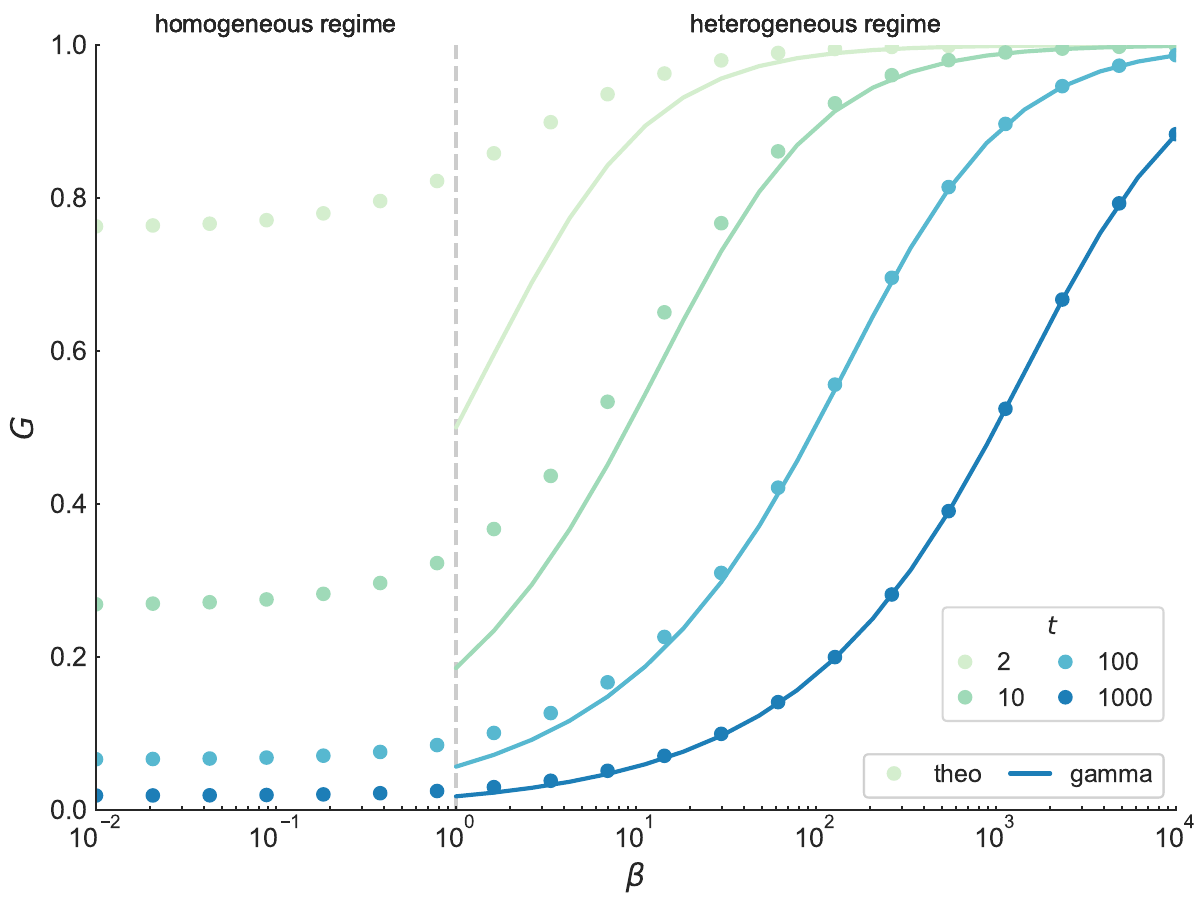}
\centering
\caption{{\bf Theoretical relation between Gini coefficient and alter preferentiality.} Gini coefficient $G$ [\eref{eq:gini_coef}] as a function of preferentiality $\beta$ in the alter activity model of Ref. \cite{iniguez2023universal}. Curves shown for several values of mean alter activity $t$, both in terms of the cumulative of the alter activity distribution $p_a$ [\eref{eq:GiniActDist}, theo], and via an approximation valid in the heterogeneous regime of $\beta \to \infty$ [\eref{eq:GiniApprox}, gamma]. Gini increases with preferentiality, indicating how alter activities become progressively more heterogeneous (i.e. unequal).}
\label{fig:gini}
\end{figure}

In the ego-net model of \cite{iniguez2023universal}, and following \esref{eq:GiniActDist}{eq:GiniApprox}, the Gini coefficient of alter activity increases monotonically with preferentiality (\fref{fig:gini}). In the homogeneous regime of random alter choice ($\beta \to 0$), $G$ is low, in consistency with a homogeneous (Poissonian) activity distribution. As $\beta$ increases and preferential alter selection becomes dominant, Gini approaches the value $G = 1$ corresponding to a heterogeneous activity distribution and complete inequality (one alter dominates all activity). The approximation of \eref{eq:GiniApprox} works well in the heterogeneous regime, but progressively fails as $\beta$ becomes small, depending on the value of the mean alter activity $t$.

\subsection{Pros and cons of each measure}
    
In the previous Sections (\ref{subsubsection:empirical_correlation} and \ref{subsubsection:theoritical_relation}), we observe that the Gini coefficient and alter preferentiality $\beta$ are closely related. Here, we review their differences and discuss the pros and cons of using each measure.

\begin{figure}
\includegraphics[scale=0.4]{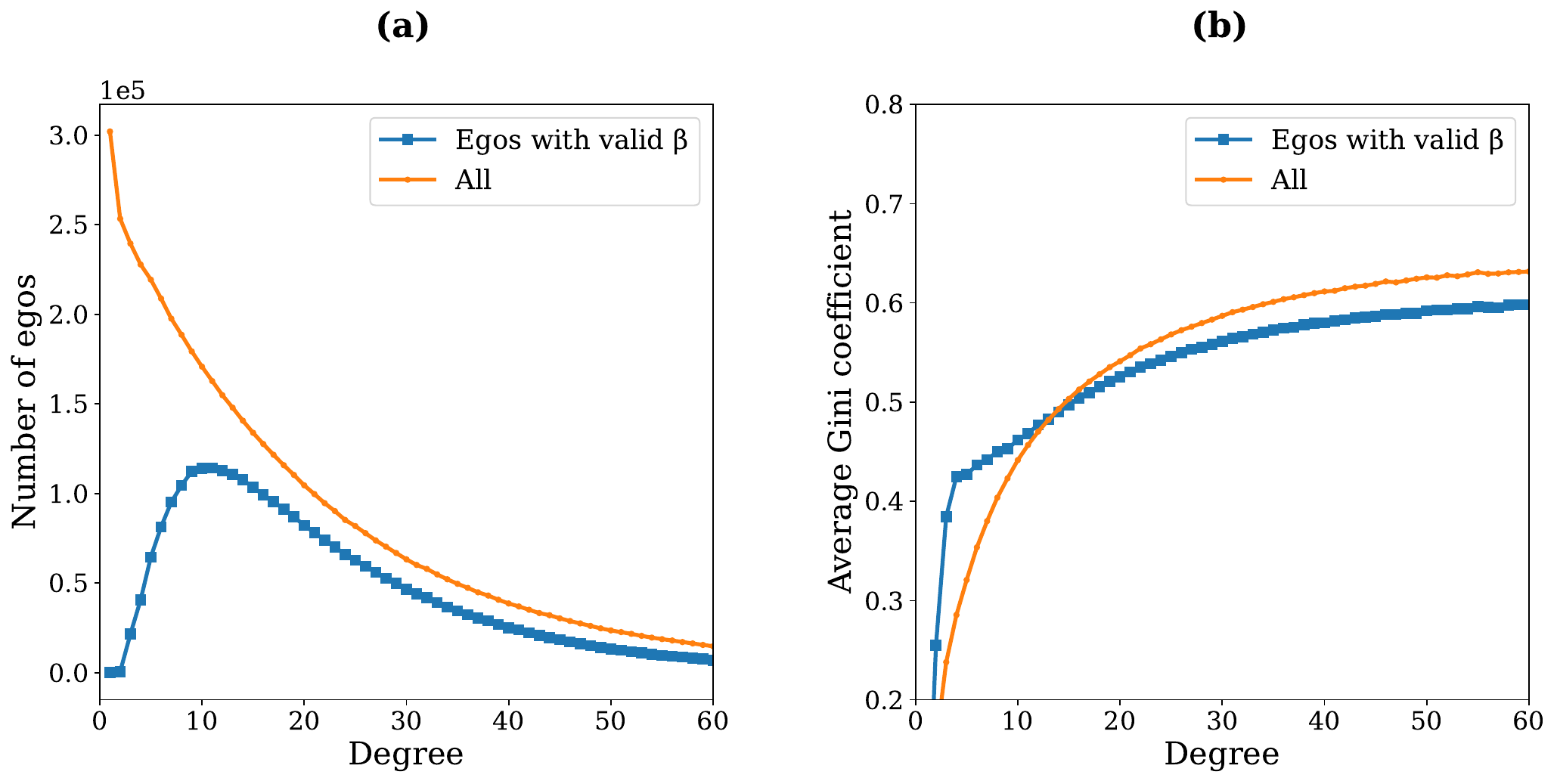}
\centering
\caption{{\bf The Gini coefficient can be calculated for any ego-net with two or more alters, while not all ego-nets can be assigned an alter preferentiality $\beta$. (left)} Total number of call ego-nets in the LMP dataset for each degree is shown by orange dots. Blue squares show the number of egos for which we can fit a $\beta$. The parameter $\beta$ can be fitted only for a small fraction of low-degree egos. {\bf (right)} Gini coefficient averaged over egos of a specific degree as a function of their degrees. The average values over all the egos and only the egos with valid $\beta$ are shown respectively in orange and blue.}
\label{fig:fraction_egos_with_valid_beta_call}
\end{figure}

The Gini coefficient can be calculated for any ego-net with two or more alters. In contrast, we might not necessarily find a valid alter preferentiality $\beta$ for all ego-nets. For instance, the model used for obtaining $\beta$ is not well defined for fully homogeneous networks where all alters have exactly the same communication amount (equivalent to a zero Gini coefficient). This prerequisite filters out those low-activity egos whose ego-nets consist of a few once-contacted alters. \fref{fig:fraction_egos_with_valid_beta_call} shows the number of call ego-nets in the LMP dataset for which we can fit $\beta$ as a function of their degrees. We observe that most of the failed cases are ego-nets with low degrees.

\section{Measuring across-channel similarity using $\Delta k$-specific references}
\label{subsection:self-similarity}

The LMP dataset used in our study is also used in \cite{heydari2018multichannel}, where the authors showed that call and SMS social signatures are not only persistent but also resemble each other for each ego. The study used Jensen-Shannon as a distance function. Here, we replicate the same study by using three other distance functions besides JSD distance and also by controlling for different levels of degree change (See Section \ref{subsection:heterogeneity_measures} for the definition of the distance functions). We want to check if the similarity of call and SMS ego-nets observed in \cite{heydari2018multichannel} is more than a side effect of people having similar degrees across these channels as compared with the population degree variation (See \fref{fig:self_vs_ref_delta_k} for evidence on the degree similarity).

To measure the self-distance for ego $e$, we construct a weighted ego-net using the call log during the whole observation window (denoted as $e_c$) and another using the SMS log (denoted as $e_s$) and use the distance functions introduced in \ref{subsection:heterogeneity_measures} to measure the heterogeneity difference of ego $e$ across the channels denoted as $d(e_c,e_s)$.

Then, we calculate the across-channel similarity values both using a single degree-agnostic reference and $|\Delta(k)|$-specific references. To construct the degree-agnostic reference, we choose a random sample of $\sim 7000$ egos and calculate the heterogeneity distance between the call ego-net of ego $x$ and the SMS ego-net of ego $y$ ($y \neq x$) for all the possible pairs in the sample. To form $|\Delta(k)|$-specific reference, we take the subset of the sampled pairs which satisfy $|k_{e,c} - k_{e,s}| = |k_{x,c} - k_{x,s}|$, where $k_{e,c}$ and $k_{e,s}$ are the degrees of ego $e$ in call and SMS communication channels, and $k_{x,c}$ and $k_{y,s}$ are the degrees of egos $x$ and $y$ respectively in channels $c$ and $S$. We define the across-channel similarity for ego $e$ by a signed dimensionless quantity defined similarly to the measure of persistence in Section \ref{subsection:persistence}
 \begin{equation}
 \label{eq:self_similarity_same_ref}
z(d(e_c,e_s), \mathrm{ref}) = \frac{d(e_c, e_s) - \mathrm{avg}(\mathrm{ref})}{\mathrm{std}(\mathrm{ref})},
\end{equation}
where $\mathrm{avg}(\mathrm{ref})$ and $\mathrm{std}(\mathrm{ref})$ denote the average and the standard deviation of the reference distribution of choice. Thus, across-channel similarity is measured as the deviation of $d(e_c, e_s)$ from the mean value of the $\mathrm{ref}$ distribution in units of standard deviation of $\mathrm{ref}$. A negative value indicates that the difference in the ego-net of ego $e$ across two channels is less than the average distance between the network of a random ego in channel $c$ and another random ego in channel $s$.

\begin{figure}[htbp]
\includegraphics[scale = 0.34]{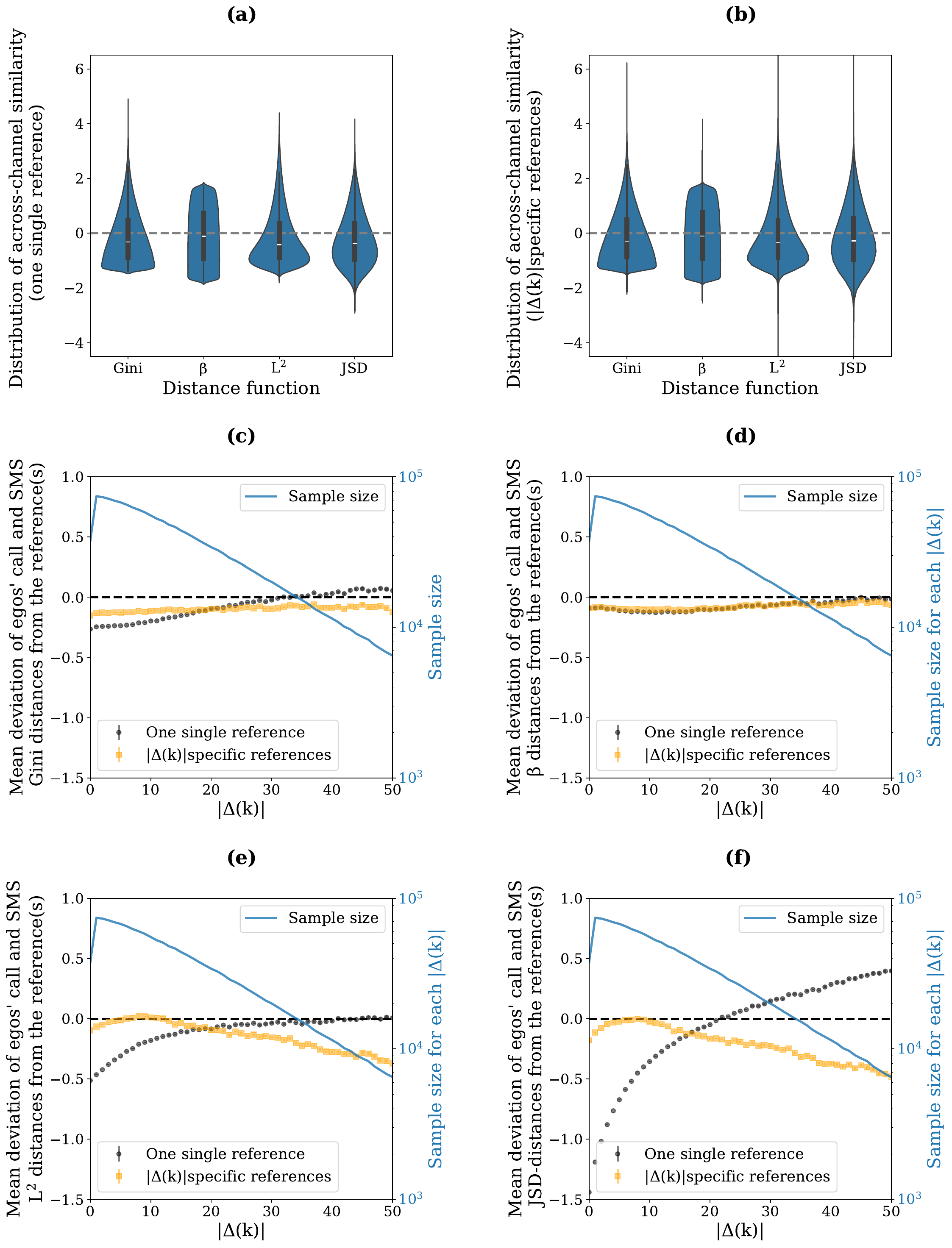}
\centering
\small
\caption{{\bf Call and SMS ego-net similarity for the egos in the LMP dataset using different measures and distance functions.} {\bf (a)} Distribution of similarity scores with a single reference. {\bf (b)} Distribution of similarity scores with $|\Delta(k)|$ specific references. In both cases and for all the measures, the majority of the scores are negative, indicating across-channel similarity. In panels {\bf (c)} to {\bf (f)} we see average call-SMS ego-net similarity scores as a function of $|\Delta(k)|$ (absolute degree difference of call and SMS ego-nets) using Gini index in {\bf (c)}; alter preferentiality $\beta$ in {\bf (d)}; $L^2$ distance in {\bf (e)}; and Jensen-Shannon distance in {\bf (f)}. 
The average across-channel similarity scores are visualized in black if they are measured with respect to one single reference distribution (regardless of $|\Delta(k)|$) and in orange if $|\Delta(k)|$-specific reference distributions are used (see Eq. \ref{eq:self_similarity_same_ref} for definition). We see that all four heterogeneity measures suggest a slight amount of across-channel similarity, at least for the lower $|\Delta(k)|$ values. However, the effect is not as strong as what we observe for the persistence of ego-nets (compared with \fref{fig:persistence}). Similarly to the case for the persistence curves, we observe that call-SMS similarity curves for the Gini coefficient and $\beta$ do not change much when we switch from a single reference to $|\Delta(k)|$-specific references. This is another evidence showing that when comparing ego-nets of different sizes, Gini and $\beta$ are better choices than $L^2$ and JSD distances. The blue curves in plots (c)-(f) show the number of egos with a particular absolute call and SMS degree-difference ($|\Delta(k)|$) (the values are on the right axes).}
\label{fig:call_sms_similarity}
\end{figure}

For all four distance functions, the across-channel similarity observation is confirmed when controlling for different levels of degree change (by using $|\Delta(k)|$-specific references). The average across-channel similarity values are negative across the $|\Delta(k)|$ range (See \fref{fig:call_sms_similarity}). However, the deviations suggesting across-channel similarity are smaller than the persistence deviations (shown in Section \ref{subsection:persistence_degree_dependency}), indicating a weaker effect. Similarly to the case for persistence measures, the across-channel similarity plots for the Gini coefficient and $\beta$ do not change much when switching from the degree-agnostic reference to $|\Delta(k)|$-specific references. This is another evidence that $\beta$ and Gini are better measures for comparing ego-nets of different sizes.

\section{Are only ego-nets with valid alter-preferentiality values persistent?}
As shown in Figure \ref{fig:fraction_egos_with_valid_beta_call}, the alter-preferentiality parameter $\beta$, which is associated with the ego-net growth model introduced in \cite{iniguez2023universal}, cannot be assigned to all the ego-nets. Particularly, the goodness-of-fit test fails for a large fraction of low-degree ego-nets. The persistence plots in Figure \ref{fig:persistence} only include the ego-nets that pass the test, because one of the objectives of our study is to compare the distance function based on $\beta$ with the other heterogeneity distance functions.

\begin{figure}[htbp]
\includegraphics[scale = 0.35]{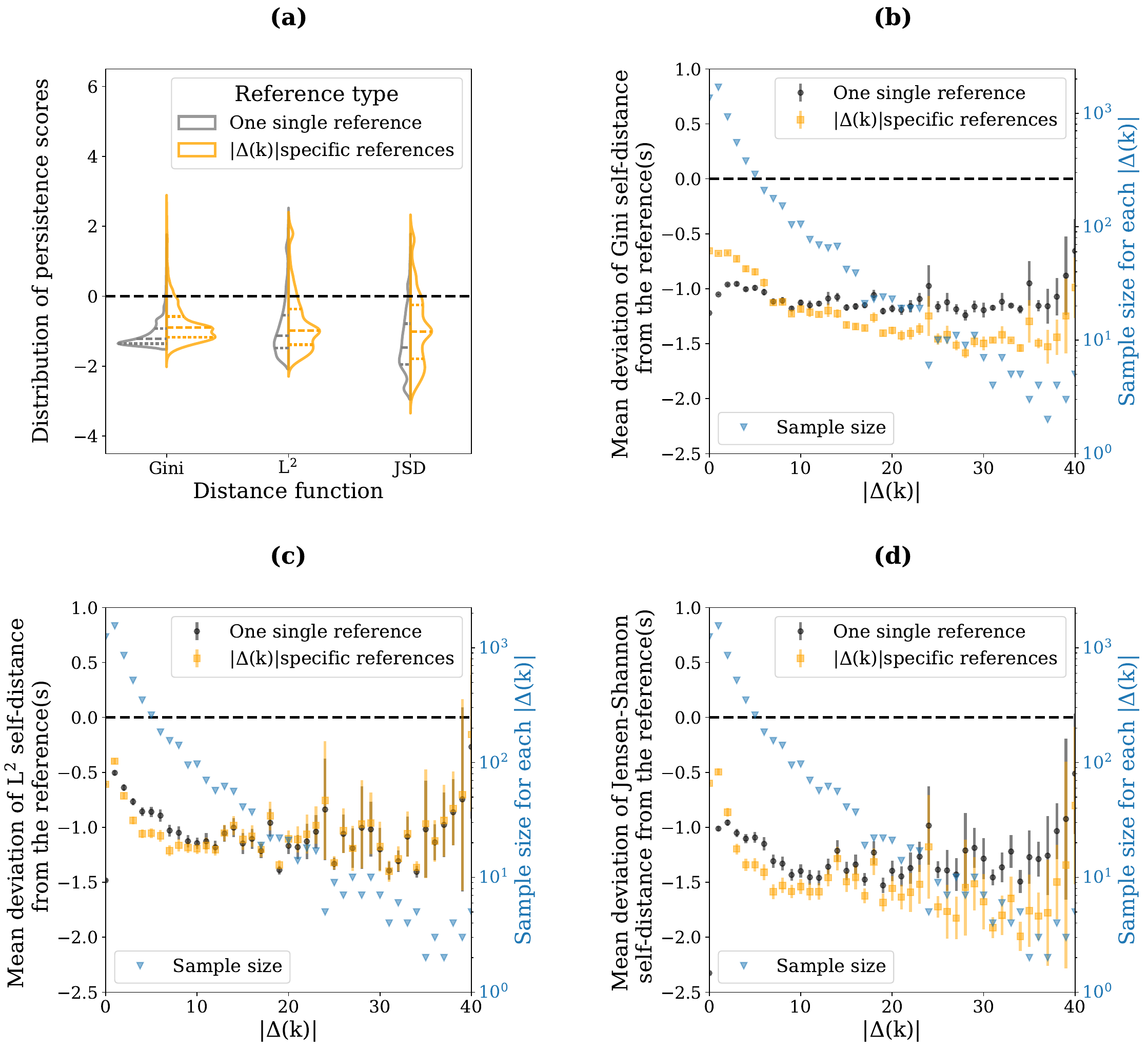}
\centering
\small
\caption{{\bf Structural persistence of call ego-nets for egos in the LMP dataset that can be assigned Gini values but not valid $\beta$.} {\bf (a)} Distribution of persistence scores with regard to a single degree-agnostic reference (grey violin-plots on the left) and with regard to $|\Delta k|$-specific references (orange violin-plots on the right). Most values for both references and all three distance functions are negative, indicating the persistence of the call ego-nets for these egos. Average persistence as a function of $|\Delta(k)|$ (absolute degree change) is shown in {\bf (b)} using the Gini distance function; in {\bf (c)} using the $L^2$ distance; and in {\bf (d)} using the JSD distance. 
The average persistence scores are visualized in grey if they are measured against one single degree-agnostic reference distribution (regardless of $|\Delta(k)|$) and in orange if against $|\Delta(k)|$-specific reference distributions. (See Eq. \ref{eq:persistence_same_ref} for the definition of the persistence score.) The error bars illustrate the standard error of the mean of the samples. All three heterogeneity measures have negative average values across the degree range. The blue triangles in plots (b)-(d) show the number of egos with each particular absolute degree change ($|\Delta(k)|$) (the values are on the right axes).
}
\label{fig:persistence_valid_gini_invalid_beta}
\end{figure}

This raises a question: Are only ego-nets with valid $\beta$s persistent? To answer this, we repeat the study using the other three distance functions for the egos that can be assigned Gini values to their call ego-nets (degree larger than one in both periods) but not valid $\beta$ (at least in one of the periods). There are 2708055 egos that fulfil this criterion in the LMP dataset. We observe that the ego-nets are persistent for these egos as compared to both the degree-agnostic and $|\Delta(k)|$-specific references (See Figure~\ref{fig:persistence_valid_gini_invalid_beta}).

\end{document}